\documentclass[12pt]{article}

  \usepackage{amsfonts}
\ifx\pdfoutput\undefined
\usepackage[dvips,bookmarks]{hyperref}
\else
\usepackage{hyperref}
\fi
\hypersetup{colorlinks=false,bookmarksopen,bookmarksnumbered,citecolor=blue,
   pdfstartview=FitH}

\usepackage[dvips]{graphicx}
\usepackage{latexsym}

\usepackage{amssymb,amsfonts,amsmath}
\usepackage{graphicx} 
\usepackage{indentfirst}

 \usepackage{bbm}
\newcommand {\cC}{{\cal C}}
\newcommand {\cD}{{\cal D}}

\newcommand {\cK}{{\cal K}}
\newcommand {\cL}{{\cal L}}
\newcommand {\cM}{{\cal M}}
\newcommand {\cN}{{\cal N}}

\newcommand {\cS}{{\cal S}}
\newcommand {\cT}{{\cal T}}
\newcommand {\cU}{{\cal U}}
\newcommand {\cV}{{\cal V}}

\newcommand{\bS}{{\bf S}}
\newcommand{\bT}{{\bf T}}

\newcommand{\bV}{{\bf V}}

\def\a{\alpha}

\def\b{\beta}

\def\d{\delta}

\def\g{\gamma}

\def\m{\mu}

\def\q{\theta}

\def\s{\sigma}

\def\D{\Delta}
\def\F{\Phi}

\def\L{\Lambda}
\def\O{\Omega}

\def\rd{{\rm d}}
\def\ri{{\rm i}}
\def\re{{\rm e}}

\newcommand{\ve}{\varepsilon}                            

\newcommand{\pa}{\partial}                           
\newcommand{\hf}{\frac12}

\newcommand{\be}{\begin{equation}}
\newcommand{\ee}{\end{equation}}
\newcommand{\bea}{\begin{eqnarray}}
\newcommand{\eea}{\end{eqnarray}}
\newcommand{\non}{\nonumber}
\newcommand{\1}{{\underline{1}}}
\newcommand{\2}{{\underline{2}}}

\newcommand{\bm}[1]{\mbox{\boldmath$#1$}}

\def\double #1{#1{\hbox{\kern-2pt $#1$}}}

\renewcommand{\Bar}{\overline}

\newcommand{\qb}{{\bar{\theta}}}

\newcommand{\bmL}{{\bm L}}
\newcommand{\bmR}{{\bm R}}
\newcommand{\bpl}{{\boxplus}}
\newcommand{\bmn}{{\boxminus}}
\newcommand{\opl}{{\oplus}}
\newcommand{\omn}{{\ominus}}

\newif\ifdtup

\def\de{{\nabla}}                                         
\def\cm{{\cal M}}

\def\deb{{\bar{\de}}}

\def\pp{+\hspace{-0.04in}+}

\newcommand{\bsubeq}{\begin{subequations}}
\newcommand{\esubeq}{\end{subequations}}

\begin{document}

\begin{titlepage}
\begin{flushright}
UMD-PP-09-064\\
December, 2009\\
\end{flushright}
\vspace{5mm}

\begin{center}
{\Large \bf On 2D $\cN=(4,4)$
superspace supergravity\footnote{Contribution to the proceedings of the Workshop 
``Supersymmetries and Quantum Symmetries'' (SQSÕ09), Bogoliubov Laboratory of Theoretical 
Physics, JINR, Dubna, July 29-August 3 2009}
}\\ 
\end{center}

\begin{center}

{\bf
Gabriele Tartaglino-Mazzucchelli\footnote{gtm@umd.edu}
} \\
\vspace{5mm}

{\it Center for String and Particle Theory,
Department of Physics, 
University of Maryland\\
College Park, MD 20742-4111, USA}
~\\

\end{center}
\vspace{5mm}

\begin{abstract}
\baselineskip=14pt

We review some recent results obtained in studying superspace formulations 
of 2D $\cN=(4,4)$ matter-coupled supergravity.
For a superspace geometry described by the minimal supergravity multiplet, 
we first describe how to reduce to components
the chiral integral by using ``ectoplasm'' superform techniques as in arXiv:0907.5264
and then we review the bi-projective superspace formalism introduced in arXiv:0911.2546.
After that, we elaborate on the curved bi-projective formalism providing a new result:
 the solution of the 
covariant type-I twisted multiplet constraints in terms of a weight-$(-1,-1)$ bi-projective superfield.

\end{abstract}
\vspace{1cm}

\vfill
\end{titlepage}

\newpage
\renewcommand{\thefootnote}{\arabic{footnote}}
\setcounter{footnote}{0}


\section{Introduction}
\setcounter{equation}{0}

In the literature,
two superspace frameworks  have been developed to study 
supersymmetric field theories with eight real supercharges.
They go under the names of harmonic superspace (HS) \cite{harm1,harm}
and projective superspace (PS) \cite{KLR,LR}.\footnote{See \cite{LRtwistor} for a review on 
flat 4D $\cN=2$ projective superspace.}
Although in some respects similar, 
the two formalisms differ in the structure of the off-shell supermultiplets 
and the supersymmetric action principle. For these reasons
 the two approaches often prove to be  complementary one
to each other.\footnote{For global supersymmetry, the relationship 
between the harmonic and projective superspace has been described
in \cite{K-double}. See also \cite{JainSiegel} for a recent discussion.}
 This proves to be confirmed when one
considers curved extensions of the HS and PS approaches.
 
An HS description of 4D $\cN=2$ conformal supergravity
was given twenty years ago \cite{Galperin:1987em}. 
This is based on a prepotential formulation 
but its  relationship to standard, curved superspace geometrical methods has not 
been elaborated in detail yet.
On the other hand,
first for five-dimensional \cite{KT-M_5D,KT-M_5Dconf} 
and then four-dimensional  \cite{KLRT-M_4D-1,KLRT-M_4D-2} 
supergravity we recently proposed a PS approach to study 
supergravity-matter systems in 
a covariant geometric 
way.\footnote{See \cite{KT-M_5DconfFlat,KT-M_4DconfFlat,Kdual,KT-M-normal} for recent 
developments and applications. Note that, the curved PS is built on the superconformal projective
multiplets of \cite{K,K2},  for a curved geometry 
projective superfields 
were first used
in studying field theory
in 5D $\cN=1$ anti-de Sitter superspace \cite{KT-M}.}
In many respect the PS formalism resembles
the covariant Wess-Zumino superspace approach to 4D $\cN = 1$ supergravity \cite{WZ-s}
even if the PS supergravity prepotential structure is still not completely understood.

Together with the formulation of general supergravity-matter systems in superspace,
one has to face the problem of reduction to components which is important for many applications.
Even if  in principle trivial, in supergravity theories, 
the components reduction of supersymmetric actions 
has always represented a challenging technical task.
At the present time, the state of the art methods are represented by superspace normal 
coordinates \cite{AD,McA,GKS,KT-Mnc}
and the so called ``ectoplasm'' \cite{Gates,GGKS,GKT-M_2009}
techniques.\footnote{We refer the reader to \cite{KT-Mnc} and \cite{GKT-M_2009}
for a more detailed list of references on normal coordinates and ectoplasm techniques.}
As described in \cite{GKS,KT-Mnc}, a crucial property of the normal coordinates approach
is its universality.
On the other hand, the ectoplasm, which is based on the use of 
superforms, is a very general method to construct locally supersymmetric invariants
\cite{Gates,GGKS}.
Moreover, in conjunction with additional ideas, the ectoplasm technique
 has proven to be flexible enough to provide 
the most efficient approach to component reduction in supergravity \cite{GKT-M_2009}.

As part of a program aimed to develop efficient off-shell superspace formulations for 
matter-coupled supergravity theories 
with eight real supercharges in various dimensions, this year we studied some
topics in the case of 2D $\cN=(4,4)$ supergravity \cite{GTM-2DSG,GatesGT-M-09-1}. 
A better understanding of locally conformal 
matter systems coupled to 2D $\cN=(4,4)$ supergravity are interesting in studying 
WZNW/Liouville-type systems, 
non-linear sigma models and $\cN=(4,4)$ non-critical strings. 
Moreover, being some aspects of 2D superspace supergravity simpler compared to D$>$2,
a better understanding of the 2D case could shed light on unclear aspects of the higher 
dimensional cases.

The main scope of this note is to review some results we recently obtained in 
\cite{GTM-2DSG,GatesGT-M-09-1}.
In particular, in \cite{GatesGT-M-09-1}, by using ectoplasm techniques, we derived
the chiral action principle in components for the case of the minimal supergravity geometry of
 Gates {\it et.~al} \cite{2DN4SG}.
 
In \cite{GTM-2DSG} the main result is represented by the formulation of a curved bi-projective
superspace for 2D $\cN=(4,4)$ conformal supergravity extending the flat case studied in
 \cite{GHR,BusLinRoc,RSS,LR-biProj}.\footnote{It is worth to note that
 for 2D $\cN=(4,4)$ supersymmetry,
harmonic superspace has been introduced in \cite{IvanovSutulin}.
A prepotential formulation for 2D $\cN=(4,4)$ conformal supergravity has been given
in the so called bi-harmonic superspace \cite{BellucciIvanov}.}
This includes the definition of a large class of matter multiplets coupled to 2D $\cN=(4,4)$ 
conformal supergravity and a manifestly locally supersymmetric and super-Weyl invariant 
action principle in bi-projective superspace.\footnote{Note that in this notes we will focus on 
the geometry given by the 
minimal supergravity multiplet of  \cite{2DN4SG} 
even if the bi-projective superfields were first defined in \cite{GTM-2DSG} 
on a new extended superspace geometry having tangent space group described by
the SO(1,1)$\times$SU(2)$_L\times$SU(2)$_R$ group.}

At the end  of the paper, we  include a new result.
Elaborating on the curved bi-projective formalism of \cite{GTM-2DSG},
we provide the solution of the covariant type-I twisted multiplet (TM-I) constraints 
\cite{Gnew2D,GHR,GatesKetov} 
in terms of a weight-$( -1, -1)$ bi-projective superfield. 
This is a new interesting development of \cite{GTM-2DSG} considering, for example, that the TM-I
is the constrained prepotential of the type-II twisted multiplet \cite{TM-II-IK,GatesKetov} 
which describes the
supergravity conformal compensator.

The paper is organized as follows. In section \ref{SUGRA} we review the superspace geometry
of the minimal multiplet of \cite{2DN4SG}.
According to \cite{GatesGT-M-09-1}, in section \ref{ecto} 
we describe how to derive the 2D $\cN=(4,4)$ superspace 
integration measure in components by using the ectoplasm technique.
Section \ref{projSuper} is devoted to a review of the bi-projective superspace formalism of 
\cite{GTM-2DSG}.
We then conclude with section \ref{TM-I_prep} which contains
the bi-projective prepotential for the covariant TM-I.


\section{2D $\cN=(4,4)$ minimal supergravity in superspace}
\label{SUGRA}
\setcounter{equation}{0}

In this section we review some aspects of the off-shell 2D $\cN=(4,4)$
 minimal supergravity multiplet first introduced in \cite{2DN4SG}.
We focus on the curved superspace geometry underlining the minimal supergravity.
For our 2D notations and conventions the reader should see \cite{GTM-2DSG}.

Consider a curved 2D $\cN=(4,4)$ superspace, which we will denote by $\cM^{2|4,4}$.
This is locally parametrized by coordinates
$z^M=(x^m,\q^{\mu \imath},\qb^\mu_\imath)$ where $m=0,1$, $\mu=+,-$ and $\imath=\1,\2$.
In the light-cone coordinates the superspace is locally parametrized by
$z^M=(x^{\pp},x^{=},\q^{+\imath},\qb^+_\imath,\q^{-\imath},\qb^-_\imath)$. 
The Grassmann variables are related one to each other by the complex conjugation rule
$(\q^{\mu \imath})^*=\qb^\mu_{\imath}$.

In \cite{2DN4SG} the tangent space group
was chosen to be SO(1,1)$\times$SU(2)$_\cV$
where $\cm$ and $\cV_{ij}$ denote the corresponding Lorentz and SU(2)$_\cV$
generators.
The covariant derivatives 
$\de_{{A}} =(\de_{{a}}, \de_{\a i},\deb_\a^i)$ 
(or $\de_{{A}} =(\de_{\pp},\de_{=}, \de_{+ i},\deb_+^i,\de_{- i},\deb_-^i)$)
of the minimal geometry are
\bea
\de_{A}&=&E_{A}
~+~\O_{A}\,\cm
~+~(\Phi_\cV)_{A}{}^{kl}\,\cV_{kl}
~.\label{CovDev-4}
\eea
Here $E_{{A}}= E_{{A}}{}^{{M}}(z) \pa_{{M}}$ is the supervielbein, 
with $\pa_{{M}}= \pa/ \pa z^{{M}}$,
$\O_{{A}}(z)$ is the Lorentz connection and
 $(\Phi_\cV)_{{A}}{}^{kl}(z)$ is the  
 ${\rm SU}(2)_\cV$ connections.
The action of the Lorentz generator on the covariant derivatives is
\begin{subequations}
\bea
&&{[} {\cal M} , \nabla_{\a  i} {]} = \hf 
(\g^3)_{\a} {}^{\b} \nabla_{\b  i}    ~,~~~
{[} {\cal M} , \deb_{\a }^i  {]} = \hf 
(\g^3)_{\a} {}^{\b} \deb_{\b}^i~,~~~
{ [}  {\cal M} , \nabla_{a}  {]} =
\ve_{ab}\nabla^b~,     
\label{Lorentz1}
\\
&&{[} {\cal M} , \nabla_{\pm  i} {]} = \pm\hf  \nabla_{\pm  i}    ~,~~~
{[} {\cal M} , \deb_{\pm }^i  {]} = \pm\hf \deb_{\pm}^i~,~~~
{ [}  {\cal M} , \nabla_{\buildrel \pp \over =}{]}
 =\pm\nabla_{\buildrel\pp \over =}
~.     
\label{Lorentz2}
\eea
\end{subequations}
The generator $\cV_{ij}$ acts on the covariant derivatives according to the rules
\bea
&&{[}  {\cV}{}_{kl} , \nabla_{\a i} {]} = \hf C_{i(k} \nabla_{\a l)} 
~,~~~
{[}  {\cV}{}_{kl} , \deb_{\a}^i  {]} = -\hf \d^{i}_{(k} \deb_{\a l)} 
~,~~~
[\cV_{kl},\de_a]=0
~.
\label{V_kl}
\eea
It is worth to note that the operator $\cV_{kl}$ generates a diagonal 
SU(2)$_\cV$ subgroup inside a SU(2)$_L\times$SU(2)$_R$ whose generators
$\bmL_{ij}$ and $\bmR_{ij}$ satisfies
\begin{subequations}
\bea
{[}  {\bmL}{}_{kl} , \nabla_{+ i} {]} = \hf C_{i(k} \nabla_{+ l)} ~,~~
{[}  {\bmL}{}_{kl} , \deb_{+}^i  {]} = -\hf \d^{i}_{(k} \deb_{+ l)} ~,~~
{[}  {\bmL}{}_{kl} , \nabla_{- i} {]} = {[}  {\bmL}{}_{kl} , \deb_{-}^{ i} {]}=0 ~,~~~~~~
\label{L_kl}
\\
{[}  {\bmR}{}_{kl} , \nabla_{- i} {]} = \hf C_{i(k} \nabla_{- l)} ~,~~
{[}  {\bmR}{}_{kl} , \deb_{-}^i  {]} = -\hf \d^{i}_{(k} \deb_{- l)} ~,~~
{[}  {\bmR}{}_{kl} , \nabla_{+ i} {]} = {[}  {\bmR}{}_{kl} , \deb_{+}^{ i} {]}=0 ~.~~~~~~
\label{R_kl}
\eea
\end{subequations}
Moreover, it holds 
$[\bmL_{kl},\de_a]=[\bmR_{kl},\de_a]=0$.
In terms of $\bmL_{ij}$ and $\bmR_{ij}$ the generator $\cV_{ij}$ is
\bea
\cV_{kl}=\bmL_{kl}+\bmR_{kl}~.
\eea
The generators $\bmL_{ij}$ and $\bmR_{ij}$ will be largely used in section \ref{projSuper}.

Note also that in \cite{GTM-2DSG} it has been formulated an extended 
supergravity multiplet whose superspace geometry is based on the
 SO(1,1)$\times$SU(2)$_L\times$SU(2)$_R$
 tangent space group. The minimal multiplet arise from the extended one after partially gauge 
 fixing the super-Weyl transformations and gauge fixing 
 the local chiral SU(2)$_\cC$ transformations generated by
 \bea
 \cC_{kl}=\bmL_{kl}-\bmR_{kl}~,~~~
 {[}  {\cC}{}_{kl} , \nabla_{\a i} {]} = \hf C_{i(k}(\g^3)_\a{}^\b \nabla_{\b l)} 
 ~.
 \eea

The minimal supergravity gauge group is given by local general coordinate and tangent space 
transformations of the form 
\bea
\d_\cK \de_{{A}} =[\cK  , \de_{A}]~,
~~~
\cK = K^{{C}} \de_{{C}} +K\cm
+(K_\cV)^{kl} \cV_{kl} 
~,
\label{SUGRAgauge}
\eea
with the gauge parameters
obeying natural reality conditions, but otherwise  arbitrary superfields. 
Given a tensor superfield $\cU(z)$, with its indices suppressed, 
it transforms as:
\bea
\d_\cK \cU = \cK\, \cU~.
\label{tensor-K}
\eea

The minimal covariant derivatives algebra has the form
\bea
{[}\de_{{A}},\de_{{B}}\}&=&T_{{A}{B}}{}^{{C}}\de_{{C}}
+R_{{A}{B}}\cm
+(R_\cV)_{{A}{B}}{}^{kl}\cV_{kl}
~,
\label{algebra-4}
\eea
where  $T_{AB}{}^C$ is the torsion, $R_{AB}$ is the Lorentz curvature
and $(R_\cV)_{AB}{}^{kl}$ is the SU(2)$_\cV$ curvature.

In \cite{2DN4SG} it was proved that the off-shell 2D $\cN=(4,4)$
minimal supergravity multiplet is described by the constraints\footnote{The algebra of covariant
derivatives here is written according to the notation of \cite{GTM-2DSG} and is equivalent to the 
one given in \cite{2DN4SG} up to trivial redefinitions of the torsion superfields.}
\begin{subequations}
\bea
\{\de_{\a i},\de_{\b j}\}&=&
-4\ri C_{ij}C_{\a\b}N\cm
+4\ri(\g^3)_{\a\b}N\cV_{ij}
~,
\label{Algebra-1.1}
\\
\{\de_{\a i},\deb_\b^j\}&=&
2\ri\d_i^j(\g^a)_{\a\b}\de_a
-4\d_i^j\Big(\ri C_{\a\b}\cT
+(\g^3)_{\a\b}\cS
\Big)\cm
\non\\
&&
+4\Big(
\ri(\g^3)_{\a\b}\cT
+C_{\a\b}\cS
\Big)\cV_{i}{}^j
~,
\label{Algebra-1.2}
\\
{[}\de_a,\de_{\b j}{]}&=&
\Big(
\ri(\g_a)_{\b}{}^{\g} \cS
+\ve_{ab}(\g^b)_{\b}{}^{\g}\cT
\Big)\de_{\g j}
-\ve_{ab}(\g^b)_{\b}{}^{\g}N\deb_{\g j}
\non\\
&&
+(\g_a)_{\b}{}^\g(\deb_{\g j}N)\cM
-\ve_{ab}(\g^b)_{\b}{}^{\g}(\deb_\g^{k}N)\cV_{jk}
~,
\label{Algebra-3/2}
\eea
\bea
{[}\de_a,\de_b{]}&=&
-\hf\ve_{ab}\Bigg(
\ri(\de^{\g k}\bar{N})\de_{\g k}
+\ri(\deb^\g_{k}N)\deb_\g^k
+\Big({\ri\over 16}{[}\de^{\a(k},\de_\a^{l)}{]}\bar{N}
-{\ri\over 16}{[}\deb^{\a(k},\deb_{\a}^{l)}{]}N
\Big)\cV_{kl}
\non\\
&&
~~~~~~~~~~
+\Big({\ri\over 4}(\g^3)^{\a\b}{[}\deb_{\a k},\deb_{\b}^k{]}N
-{\ri\over 4}(\g^3)^{\a\b}{[}\de_{\a k},\de_{\b}^k{]}\bar{N}
\Big)\cm
\non\\
&&
~~~~~~~~~~
+\Big(8\cT^2
+8\cS^2
+8\bar{N}N
\Big)\cm
\Bigg)
~.~~~~~~
\label{Algebra-2}
\eea
\end{subequations}
Here the dimension-1 components of the torsion obey  the reality conditions
\bea
&(N)^*=\bar{N}~,~~~
(\cT)^*=\cT~,~~~
(\cS)^*=\cS~.
\eea
The $N$, $\cS$ and $\cT$ superfields are Lorentz scalars and are invariant under SU(2)$_\cV$
transformations.

The components of the dimension-1 torsion obey differential constraints imposed by the Bianchi 
identities.
At dimension-3/2 the Bianchi identities give
\bea
\de_{\a i}N=0
~,~~~
\de_{\a}^{i}\cS
=
{\ri\over 2}(\g^3)_\a{}^{\b}\deb_\b^{i}N
~,~~~
\de_{\a}^i\cT
=
-\hf\deb_\a^{i}N
~.
\label{minimalCONSTR}
\eea

We conclude this section by noting that, besides the SO(1,1)$\times$SU(2)$_\cV$ tangent space 
group 
transformations, the minimal supergravity multiplet provides a representation of the 
superconformal 
group through local super-Weyl transformations. This is completely analogue to the 
analysis of Howe and Tucker \cite{HoweTucker}:
super-Weyl transformations are ``scale'' variations of the covariant derivatives
 such that the torsion constraints remain invariant.
In the case of the 2D $\cN=(4,4)$ minimal supergravity multiplet,
the super-Weyl transformations are generated by
two real superfields $\bS,\,\bS_{ij}=\bS_{ji}$, $(\bS)^*=\bS$, $(\bS_{ij})^*=\bS^{ij}$,
through the following infinitesimal 
variation of the spinor covariant derivative \cite{2DN4SG,GTM-2DSG}
\bea
\tilde{\d} \de_{\a i}&=&
\hf\bS\de_{\a i}
+(\g^3)_\a{}^\b\bS_i{}^{j}\de_{\b j}
 +(\g^3)_\a{}^\g(\de_{\g i}\bS)\cM
 +(\de_{\a}^{k}\bS)\cV_{ik}
~.
\label{SW-000}
\eea
The first term in the previous equation is a local super-scale transformation while
the second term is related to a compensating chiral SU(2)$_{\cC}$ transformation of the covariant 
derivatives \cite{GTM-2DSG}.
The $\bS$ and $\bS_{ij}$ superfields have to satisfy the differential constraint
\bea
(\de_{\a i}\bS_{kl})=-\hf(\g^3)_\a{}^\b C_{i(k}(\de_{\b k)}\bS)~.
\label{dbSbS}
\eea
This is the dimension-1/2 differential constraint of a twisted-II multiplet 
\cite{TM-II-IK,GatesKetov}.

To ensure the invariance of the supergravity constraints,
the dimension-1 torsion components of the minimal multiplet have to transform according to
 the following rules \cite{GTM-2DSG}\bsubeq\bea
\tilde{\d} N&=&\bS N+{\ri\over 8} (\g^3)^{\g\d}(\de_{\g k}\de_{\d}^k\bS)
~,
\label{minimalSW-1}
\\
\tilde{\d}\cT&=&\bS\cT +{\ri\over 16}(\g^3)^{\g\d}({[}\de_{\g k},\deb_{\d}^k{]}\bS)
~,
\label{minimalSW-2}
\\
\tilde{\d}\cS&=&\bS\cS+{1\over 16}({[}\de_{\g k},\deb^{\g k}{]}\bS)
~.
\label{minimalSW-3}
\eea
\esubeq
The transformations of the $\deb_\a^i$ covariant derivative can be trivially obtained by 
complex conjugation of (\ref{SW-000}) while for the vector covariant derivative it holds
\bea
\tilde{\d}\de_a&=&
\bS\de_a
+{\ri\over 2}(\g_a)^{\g\d}(\de_{\g k}\bS)\deb_{\d}^k
 +{\ri\over 2}(\g_a)^{\g\d}(\deb_{\g}^{k}\bS)\de_{\d k}
 \non\\
 &&
+\ve_{ab}(\de^b\bS)\cM
- \ve_{ab}(\de^b\bS^{kl})\cV_{kl}
~.
\eea


\section{Ectoplasm and 2D $\cN=(4,4)$ superspace integration}
\setcounter{equation}{0}
\label{ecto}

The aim of this section is to review the results of \cite{GatesGT-M-09-1} 
about the component reduction of the chiral integral in 2D $\cN=(4,4)$ minimal supergravity.
According to the ectoplasm paradigm for component reduction of superspace actions
in supergravity, the search of supersymmetric invariants is related to the study of closed 
superforms \cite{Gates}.\footnote{Note that a mathematical construction giving the 
formal, but physically uncomplete, bases for 
the ectoplasm methods can be found in the theory of integration over surfaces in 
supermanifolds developed in \cite{GKSchwarz,BS,Vor}.}
Before the description of the results in \cite{GatesGT-M-09-1}, let us give a briefly review of
the ectoplasmic construction of supersymmetric actions.

Consider a curved superspace $ \cM^{d|\d}$ with $d$ space-time 
and $\d$ fermionic dimensions, and  let $\cM^{d|\d}$ be parametrized by local coordinates 
$z^M =(x^{\hat m} , \q^{\hat \m})$, where ${\hat m} =1, \dots, d$ and ${\hat \m}=1, \dots, \d$.
The corresponding superspace geometry is described by covariant derivatives 
\bea
\de_A =(\de_{\hat a} , \de_{\hat \a})
=E_A +\F_A~, \quad 
E_A : = E_A{}^M\, \pa_M ~, \quad 
\F_A := \F_A  {\bm \cdot} {\mathbb J} =E_A{}^M \F_M ~.
\label{cov-der}
\eea
Here $\mathbb J$ denotes 
the generators of the structure group
(with all indices of $\mathbb J$s  suppressed),  
$E_A $ is the inverse vielbein, and $\F ={\rm d} z^M \F_M =E^A \F_A$ the connection. 
The vielbein $E^A := {\rm d}z^M E_M{}^A$
and its inverse  $E_A $ 
are such that
$E_A{}^M E_M{}^B =\d_A{}^B$ and
$E_M{}^A E_A{}^N =\d_M{}^N$. 
The covariant derivatives obey the algebra 
\bea
[\de_A , \de_B \} = T_{AB}{}^C \de_C + R_{AB}{\bm \cdot} {\mathbb J} ~, 
\label{torsion}
\eea
with $T_{AB}{}^C $ the torsion, and $R_{AB}$ the curvature.

Next, consider a super $d$-form 
\be
J= \frac{1}{d!} {\rm d} z^{M_d}  \wedge \dots \wedge {\rm d}z^{M_1} J_{M_1\dots M_d}=
\frac{1}{d!} E^{A_d} \wedge \dots \wedge E^{A_1}  J_{A_1 \dots A_d}
\ee
constrained to be  closed  
\be
{\rm d} J =0~\qquad \Longleftrightarrow \qquad
\de_{[B}J_{A_1\cdots A_d \}}-\frac{d}{ 2}T_{[BA_1|}{}^C J_{C|A_2\cdots A_d\}}=0~.
\label{cohomology}
\ee
Then, consider the following integral over the bosonic space-time coordinates
\bea
S&= &\frac{1}{d!} \int\rd^dx\, \ve^{{\hat m}_1 \dots {\hat m}_d }J_{{\hat m}_1 \dots {\hat m}_d}
=  \frac{1}{d!}\int\rd^d x\, \ve^{ {\hat m}_1 \dots {\hat m}_d}E_{{\hat m}_d}{}^{A_d} \dots E_{{\hat m}_1}{}^{A_1}
J_{A_1 \dots A_d}
~.
\label{ectoplasm}
\eea
Due to the closure of the super $d$-form $J$, the functional $S$ turns out to be such that:
(i) $S$ is independent of the Grassmann variables
$\q$'s; and (ii) $S$ is invariant under general coordinate transformations on $ \cM^{d|\d}$ 
and structure group transformations. 
Now, define the component vielbein as 
$e_{\hat{m}}{}^{\hat{a}}=E_{\hat{m}}{}^{\hat{a}}\big|_{\q=0}$
where its inverse $e_{\hat{a}}{}^{\hat{m}}$ is such that
$e_{\hat{m}}{}^{\hat{b}}e_{\hat{b}}{}^{\hat{n}}=\d_{\hat{m}}^{\hat{n}}$,
$e_{\hat{a}}{}^{\hat{n}}e_{\hat{n}}{}^{\hat{b}}=\d_{\hat{a}}^{\hat{b}}$.
If one defines the gravitini fields according to 
$\Psi_{\hat{a}}{}^{\hat{\a}}:=-e_{\hat{a}}{}^{\hat{m}}E_{\hat{m}}{}^{\hat{\a}}\big|_{\q=0}$
the functional (\ref{ectoplasm}) can be rewritten as
\bsubeq\bea
S&= &
\frac{1}{d!}\int\rd^d x\, \ve^{{\hat m}_1 \dots {\hat m}_d}E_{{\hat m}_d}{}^{A_d} \dots E_{{\hat m}_1}{}^{A_1}
J_{A_1 \dots A_d} \Big|_{\q=0}~,
\label{ectoplasm1}
\\
&= &
\frac{1}{d!}\int\rd^d x\,\re^{-1}\, \ve^{{\hat a}_1 \dots {\hat a}_d}\Big(
J_{\hat{a}_1\dots \hat{a}_d} 
-d\,\Psi_{{\hat a}_1}{}^{\hat{\a}}J_{\hat{\a}\hat{a}_2 \dots \hat{a}_d} 
+{d(d-1)\over 2}\,\Psi_{\hat{a}_2}{}^{\hat{\a}_2}\Psi_{{\hat a}_1}{}^{\hat{\a}_1}
J_{\hat{\a}_1\hat{\a}_2\hat{a}_3 \dots \hat{a}_d} 
\non\\
&&~~~~~~~~~~~~~~~~~~~~~~~~~~~
+\cdots
+(-)^d\Psi_{{\hat a}_d}{}^{\hat{\a}_d} \dots \Psi_{{\hat a}_1}{}^{\hat{\a}_1}
J_{\hat{\a}_1 \dots \hat{\a}_d}
\Big)\Big|_{\q=0}
~,
\label{ectoplasm2}
\eea
\esubeq
where
$\re^{-1}=[\det{e_{\hat{a}}{}^{\hat{m}}}]^{-1}$.
Besides the closure condition (\ref{cohomology}), depending on the case under consideration,
 the superform $J$ obeys some additional 
covariant constraints imposed on its components $J_{A_1 \dots A_d}$.
In cases related to component reductions of superspace actions, 
the components $J_{A_1 \dots A_d}$
are all function of a single superfield $\cL$, spinor covariant derivatives of it and 
torsion components.
The maximum number of derivatives of $\cL$ in a given component 
$J_{A_1 \dots A_d}$ depends on its mass dimension.
The cohomology equation (\ref{cohomology}) iteratively define the $J_{A_1 \dots A_d}$
 components with higher dimension in terms of derivatives and torsion multiplying
 the lower dimensional  components.


Let us now consider the case of 2D $\cN=(4,4)$ minimal supergravity.
On general grounds we can easily construct a locally supersymmetric invariant 
as
\bea
S=\int \rd^2x\,\rd^4\q\,\rd^4\qb\,E^{-1}\,\cL~,~~~~~~~~~
E^{-1}:=[{\rm Ber}\, E_{A}{}^M]^{-1}~,
\label{S}
\eea
where $\cL$ is a scalar and SU(2)-invariant but otherwise unconstrained superfield.

For practical application one is  interested to have the previous action principle ready  for 
components reduction. In particular we want to find two fourth order differential operators
$\D^{(4)}$ and $\overline{\cD}^{(4)}$ such that
\bea
S=
 \hf\int \rd^2x\, 
\,  {\rm e}{}^{-1}  \, {\D}{}^{(4)}  \, { {\cal {\Bar D}}{}^{(4)} }  \,
{\cal L}\,\Big|_{\q=0}~.
\label{Scomp}
\eea
Here with $\F|_{\q=0}$ we indicate the limit where all the Grassman variables in a superfield $\F$
are set to zero.
In (\ref{Scomp}) the operator  $\overline{\cD}^{(4)}$ defined by
\bea
 {\cal {\Bar D}}{}^{(4)} ~=~ 
 \Big({\bar {\nabla}}{}^{(2)}{}^{\a\b} 
 + 4 \ri \bar{N} \, (\g^3){}^{\a\b}\Big){\bar {\nabla}}{}^{(2)}_{\a\b} 
~,\label{ChRLprj}  
\eea
is the chiral projection operator satisfying 
\bea 
{\bar \nabla}{}_{\g}^{ i} \,   {\cal {\Bar D}}{}^{(4)}   \, \Psi 
~=~ {\bar \nabla}{}_{\g}^{ i} \,  \Big( {\bar {\nabla}}{}^{(2)}{}^{\a\b}  + 4 \ri \bar{N} \, 
(\g^3){}^{\a\b} \, \Big) {\bar {\nabla}}{}^{(2)}_{\a\b} \, \Psi  ~=~ 0
 \label{ChRLcond}   
 \eea
for any general scalar and SU(2)-invariant superfield $ \Psi$. 
Here the operator $\deb^{(2)}_{\a\b}$ is
 \bea
{\bar \nabla}{}^{(2)}_{\a\b}  &=&  \hf \, C_{ij}\Big(
{\bar \nabla}_{\a}^{ i}{\bar \nabla}_{\b}^{j}
+{\bar \nabla}_{\b}^{ i}{\bar \nabla}_{\a}^{j} \Big)
~.
\eea
The chiral projector (\ref{ChRLprj}) for the minimal supergravity was recently computed in 
\cite{GatesMorrison} by Gates and Morrison.

The operator $\D^{(4)}$ is called the ``chiral'' density projector operator. We computed it in 
\cite{GatesGT-M-09-1} by using ectoplasm techniques.
The fact that it exist the factorization ${\D}{}^{(4)}  \, { {\cal {\Bar D}}{}^{(4)} } $ in (\ref{Scomp})
is due to the existance of covariantly chiral superfield and integration over the chiral subspace 
for the 2D $\cN=(4,4)$ minimal supergravity.\footnote{One expects similar factorizations every time
invariant subspaces of a given curved superspace exist.}
In the ectoplasm framework the factorization results trivial.

The construction of the density projector operator using ectoplasm lies in the existence of a 
``chiral''  closed two form which is function of an unconstrained covariantly chiral superfield
$U$ such that $\deb_\a^i U=0$.
The chiral superfield $U$ plays the role of the chiral Lagrangian and can be thought as
$\Bar{\cD}^{(4)}\cL$ by using the chiral projector.

The components  
$  J_{A}{}_{B} = \big(J_{\a i}{}_{\b j} ,   J_{\a  i}{}_{\b}^{j} ,  J_{\a}^i{}_{\b}^j , 
J_{\a i}{}_{b},  J_{\a}^i{}_{b} , J_{ab}\big)$
 of the closed super two-form we are interested in turn out to be
 \bsubeq
  \bea
 J_{\a}^i{}_{\b}^j &=&
\Big( 2 (\g^3)_{\a\b}\de^{(2)}{}^{ij}
-C_{\a\b}C^{ij}(\g^3)^{\g\d}\de^{(2)}_{\g\d}\Big){U}
~,
\label{2-form-a}
\\
J_{\a}^i{}_{b}
&=& -
\frac{\ri}{3} \ve_{bc}(\g^c)_{\a}{}^{\g}\de_{\g k}\de^{(2)}{}^{ik}{U}
~,
\label{2-form-b}
\\
J_{ab}
&=&
-\frac{1}{8} \ve_{ab}\, \Big(
\de^{(4)} +4\ri N(\g^3)^{\a\b}\de^{(2)}_{\a\b}\Big){U}
~,
\label{2-form-c}
\\ 
J_{\a i}{}_{\b j}&=& 
J_{\a  i}{}_{\b}^{j}~=~
J_{\a i}{}_{b}~=~0
~.
\label{2-form-d}
  \eea
  \esubeq
Here we have introduced second and fourth order spinorial derivative
operators via the equations 
 \bea
\nabla{}^{(2)}_{\a\b} =  \hf
\big(\nabla_{\a k}\nabla_{\b}^k +\nabla_{\b k}\nabla_{\a}^k\big) 
~, ~~
\nabla{}^{(2)}_{ij} = \hf \big(
\nabla_{\g i} \nabla^\g_{j}  + \nabla_{\g j} \nabla^\g_{i}\big)
~,~~
\nabla{}^{(4)}  =\frac{1}{3}\nabla{}^{(2)}{}^{kl} \nabla{}^{(2)}_{kl}~.~~~
\label{DiffOps}  
\eea

The complex closed super two-form (\ref{2-form-a})--(\ref{2-form-d}) satisfies the equation 
(\ref{cohomology}) where the supergravity geometry is the 2D $\cN=(4,4)$ minimal one of
section \ref{SUGRA}.
A way to derive (\ref{2-form-a})--(\ref{2-form-d}) is to first take the following ansatz for the lower 
dimensional components\footnote{In \cite{GatesGT-M-09-1} we derived a real closed super 
two-form which is function of $U$ and its antichiral complex  conjugate $\bar{U}$.
It is easy to observe that the chiral and antichiral sectors are algebraically independent 
under eq. (\ref{cohomology}). Then, relaxing the reality condition, one can find the closed super 
two-form (\ref{2-form-a})--(\ref{2-form-d}) with computations equal to the one given in 
\cite{GatesGT-M-09-1}; the result is in fact identical but with the antichiral sector formally 
turned off.}: 
$J_{\a i}{}_{\b j}=J_{\a  i}{}_{\b}^{j}=0$, 
$
J_{\a}^i{}_{\b}^{j}=
\big(a(\g^3)_{\a\b}\de^{(2)}_{ij}
+b\,C_{\a\b}C_{ij}(\g^3)^{\g\d}\de^{(2)}_{\g\d}
+C_{\a\b}C_{ij}F
\big)U
$
where
$
F=\big(b_1N+b_2\bar{N}+b_3\cS+b_4\cT\big)
$
and $a,\, b,\, b_1,\,b_2,\,b_3,\,b_4$ are constants.
Imposing the closure equation (\ref{cohomology}) on the components $J_{AB}$ one 
fixes the constants
and iteratively express the higher dimensional components $J_{\a i}{}_b,\,J_{\a}^i{}_b$ and 
$J_{ab}$ in terms of derivatives of the lower dimensional one. This procedure gives
the result (\ref{2-form-a})--(\ref{2-form-d}).

It would be interesting to re-derive the previous closed super two-form by using the powerful 
arguments recently developed in \cite{GKT-M_2009} and in particular find a 2D
$\cN=(4,4)$ ``chiral'' closed super 1-form such that from its square wedge product one can derive
the closed 2-form just introduced.


To conclude let us give the component form of the action (\ref{Scomp}) by using the ectoplasm 
functional (\ref{ectoplasm2}).
In the 2D $\cN=(4,4)$ case, eq. (\ref{ectoplasm2}) becomes
\bea
S&=&\hf \int \rd^{2} x  \, {\rm e}^{-1} \,  \ve^{ab}\Big(
{J}_{ ab } 
-2\big(
{\bar \psi}_{a} {}^{ \a}_i {J}_{ \a}^i{}_{b }
+\psi_{ a}{}^{\a i} {J}_{\a i}{}_{b}\big)
-{\bar \psi}_{a} {}^{ \a}_i{\bar \psi}_{b} {}^{ \b}_j {J}_{\a}^i{}_{  \b  }^j 
\non\\
&&~~~~~~~~~~~~~~~~~~~~~~
-2\psi_{ a} {}^{\a i}{\bar \psi}_{ b} {}^{ \b}_j {J}_{\a}{}_i {}_{ \b}^j 
-\psi_{a} {}^{\a i } \psi_{ b}{}^{\b j}  {J}_{\a i}{}_{\b j} 
\Big){\Big |}_{\q=0}  
~.
\label{IntTheory4}
\eea
By using the previous expression, the equations (\ref{2-form-a})--(\ref{2-form-d}) and
the chiral superfield $U=\Bar{\cD}^{(4)}\cL$,
one finds the component action (\ref{Scomp}) to be
\bea
S&=& \int \rd^{2} x \, {\rm e}^{-1} \, {\Big(} 
  \frac{1}{8} \de^{(4)} + \frac{\ri}{2} N (\g^3)^{\a\b}\de^{(2)}_{\a\b}
+   \frac{\ri}{3}{\bar \psi}_{a} {}^{ \g}_i ( \g^a)_{\g}{}^{\d}\de_{\d j}\de^{(2)}{}^{i j}
\non\\
&&
~~~
-   \ve^{ab} {\bar \psi}_{a} {}^{ \a}_i {\bar \psi}_{b} {}^{ \b}_j (\g^3)_{\a\b}\de^{(2)}{}^{ij} 
-\frac{1}{2}\ve^{ab} {\bar \psi}_{a} {}^{ \a}_i
{\bar \psi}_{b}{}_{\a}^i(\g^3)^{\g\d}\de^{(2)}_{\g\d}
{\Big)} 
{\Bar{\cD}}^{(4)}{\cal L}  {\Big |}_{\q=0} 
~.
 \label{IntTheory5}  
 \eea
The terms in the brackets then define the ``chiral'' density projector operator $\D^{(4)}$.


\section{Curved bi-projective superspace}
\setcounter{equation}{0}
\label{projSuper}

In section \ref{SUGRA} we have reviewed the geometric description 
of 2D $\cN=(4,4)$ minimal supergravity in superspace \cite{2DN4SG}.
Let us now turn to discuss a large family of off-shell supermultiplets
coupled to supergravity, which can be used to describe supersymmetric matter.
We introduced them in \cite{GTM-2DSG} 
under the name of covariant bi-projective supermultiplets. 
These supermultiplets are a curved-superspace extension of the 2D 
multiplets introduced in the flat case in \cite{BusLinRoc,GHR,RSS,LR-biProj}.
The formalism possesses clear similarities with the bi-harmonic superspace approach of 
\cite{IvanovSutulin,BellucciIvanov}.
Moreover,
curved bi-projective superspace is a 2D extension of the curved projective approach
 recently developed in the cases of 5D $\cN=1$ supergravity \cite{KT-M_5D,KT-M_5Dconf} 
 and 4D $\cN=2$ supergravity \cite{KLRT-M_4D-1,KLRT-M_4D-2}.

It is useful to introduce auxiliary isotwistors coordinates $u^{\opl}_i\in 
{\mathbb C}^2\setminus \{0\}$ and $v^{\bpl}_i\in {\mathbb C}^2\setminus\{0\}$ in addition to the
superspace coordinates $z^M=(x^m,\q^{\mu \imath},\qb^\mu_\imath)$.
All the coordinates $u^{\opl}_i$, $v^{\bpl}_i$ and $z^M$ are
 defined to be inert under the action of the structure group.

The next step is to introduce superfields which are functions of $z^M$ and also of the extra 
$u^{\opl}$ and $v^\bpl$ variables and have well defined supergravity gauge transformations.
We define a weight-(m,n) bi-isotwistor superfield $U^{(m,n)}(z,u^\opl,v^\bpl)$
to be holomorphic on an open domain of 
$\{{\mathbb C}^2 \setminus  \{0\}\}\times\{{\mathbb C}^2 \setminus  \{0\}\}$ 
 with respect to 
the homogeneous coordinates $(u^{\opl}_i,v^\bpl_j) $  
for ${\mathbb C}P^1\times {\mathbb C}P^1$,
and is characterized by the  conditions:\\
(i) it is  a homogeneous function of $(u^\opl,v^\bpl)$ 
of degree $(m,n)$, that is,  
\bsubeq
\bea
U^{(m,n)}(z,c_L\,u^\opl,v^\bpl)&=&(c_L)^m\,U^{(m,n)}(z,u^\opl,v^\bpl)~, 
\qquad c_L\in \mathbb{C}\setminus \{ 0 \}~,
\label{weight-m}
\\
U^{(m,n)}(z,u^\opl,c_R\,v^\bpl)&=&(c_R)^n\,U^{(m,n)}(z,u^\opl,v^\bpl)~, 
\qquad c_R\in \mathbb{C}\setminus \{ 0 \}~;
\label{weight-n}
\eea
\esubeq
(ii)  the minimal supergravity gauge transformations act on $U^{(m,n)}$ 
as follows (remember that $\cV_{ij}=(\bmL_{ij}+\bmR_{ij})$):
\bsubeq
\bea
\d_\cK U^{(m,n)} 
&=& \Big(K^{{C}} \de_{{C}} +K\cm
+(K_\cV)^{kl}{\cV}_{kl} \Big)U^{(m,n)} ~,  
\label{localU}
\\
{\bm L}_{kl}U^{(m,n)}&=& -\frac{1}{2(u^\opl u^\omn)} \Big(u^\opl_{(k}u^\opl_{l)} D^{\omn\omn} 
-m \, u^\opl_{(k}u^\omn_{l)} \Big) U^{(m,n)} ~,
\label{LU}
 \\ 
{\bm R}_{kl}U^{(m,n)}&=& -\frac{1}{2(v^\bpl v^\bmn)} \Big(v^\bpl_{(k}v^\bpl_{l)} D^{\bmn\bmn} 
-n \, v^\bpl_{(k}v^\bmn_{l)} \Big) U^{(m,n)} ~,
\label{RU}
\\
\cM U^{(m,n)}&=&{m-n\over 2} U^{(m,n)}~.
\label{LorentzU}
\eea 
\esubeq
where we have introduced
\bsubeq
\bea
D^{\omn\omn}=u^{\omn i}{\pa\over \pa u^{\opl i}}~&,&~~~
D^{\bmn\bmn}=v^{\bmn i}{\pa\over \pa v^{\bpl i}}~,
\\
(u^{\opl}u^\omn):=u^{\opl i}u^\omn_i\ne 0~&,&~~~
(v^\bpl v^\bmn):=v^{\bpl i} v^\bmn_i \ne0~.
\label{uuvv}
\eea
\esubeq
The previous equations involve two new isotwistors  $u^\omn$ and $v^\bmn$
which are subject to the only conditions (\ref{uuvv}) and are otherwise completely arbitrary.
One can prove that, due to (\ref{weight-m}),
the superfield $(\bmL_{kl}U^{(m,n)})$ is independent 
of $u^\omn$ even if the transformations in (\ref{LU}) explicitly depend on it; 
similarly $({\bm R}_{kl}U^{(m,n)})$ is independent of $v^\bmn$. 
Then $\cV_{kl}U^{(m,n)}$ 
and, in particular, $\d_{\cK}U^{(m,n)}$ are independent of $u^\omn$ and $v^\bmn$.
One can prove that the homogeneity condition is closely related to (\ref{LU})--(\ref{RU})
and the independence on $u^\omn$ and $v^\bmn$.
The reader should see \cite{KLRT-M_4D-1} 
for a more detailed discussion on the 
SU(2) transformations of isotwistor-like superfields.
Note that, even if the supergravity gauge group of the minimal multiplet possesses only 
SU(2)$_\cV$ transformations in (\ref{localU}), it is useful to keep manifest the SU(2)$_L$
and SU(2)$_R$ parts \cite{GTM-2DSG}.

Using the $u,v$ isotwistors one can define the covariant derivatives
\bea
\de_+^\opl:=u^\opl_i\de_+^i~,~~~
\deb_+^\opl:=u^\opl_i\deb_+^i~,~~~~~~
\de_-^\bpl:=v^\bpl_i\de_-^i~,~~~
\deb_-^\bpl:=v^\bpl_i\deb_-^i~.
\eea
A crucial property of 2D bi-isotwistor superfields is that the anticommutator 
 of any of the covariant derivatives 
$\de_+^\opl,\,\deb_+^\opl,\,\de_-^\bpl,\,\deb_-^\bpl$ is zero when acting on $U^{(m,n)}$.
It holds
\bea
0=\{\de_+^\opl,\de_+^\opl\}U^{(m,n)}
=\{\de_+^\opl,\deb_+^\opl\}U^{(m,n)}
=\{\de_+^\opl,\de_-^\bpl\}U^{(m,n)}
=\cdots~.~~~~~~
\label{dedeU}
\eea
The proof of this important relation is given in \cite{GTM-2DSG}.
With the definitions (i) and (ii) assumed, the set of bi-isotwistor superfields results to be
 closed under the 
product of superfields and the action of the $\de_+^\opl,\,\deb_+^\opl,\,\de_-^\bpl,\,\deb_-^\bpl$
derivatives.
In fact, given a weight-(m,n) $U^{(m,n)}$ 
and a weight-(p,q)  $U^{(p,q)}$ bi-isotwistor superfields 
 the superfield $(U^{(m,n)}U^{(p,q)})$ is a 
weight-(m+p,n+q) bi-isotwistor superfield.
Moreover, the superfields $(\de_+^\opl U^{(m,n)})$, $(\deb_+^\opl U^{(m,n)})$ and 
$(\de_-^\bpl U^{(m,n)})$, $(\deb_-^\bpl U^{(m,n)})$ are respectively weight-(m+1,n) and 
weight-(m,n+1) bi-isotwistor superfields.

Due to (\ref{dedeU}), one can consistently define analyticity constraints. Let us then
introduce 2D $\cN=(4,4)$ covariant bi-projective superfields.
We define a weight-(m,n) covariant bi-projective supermultiplet 
$Q^{(m,n)}(z,u^\opl,v^\bpl)$ to be a bi-isotwistor superfield satisfying (i), (ii), 
(\ref{weight-m})--(\ref{LorentzU})
and to be constrained by the analyticity  conditions
\be
\de^\opl_{+} Q^{(m,n)}  =\deb^\opl_{+} Q^{(m,n)}=0~,~~~~~~
\de^\bpl_{-} Q^{(m,n)}  =\deb^\bpl_{-} Q^{(m,n)}=0~.
\label{analyticity}
\ee  
The consistency of the previous constraints is indeed guaranteed by eq. (\ref{dedeU}).

For the coupling to conformal supergraity, it is important  to derive consistent super-Weyl 
transformations of the matter multiplets.
One can prove that the transformation (remember that $\cC_{ij}=(\bmL_{ij}-\bmR_{ij})$)
\bea
 \tilde{\d}Q^{(m,n)}=\Big(\frac{m+n}{2}\bS  -\bS^{kl}\cC_{kl}\Big)Q^{(m,n)}
 ~,
 \label{projSuperWeyl}
 \eea
preserve the analyticity conditions (\ref{analyticity}).
Note the presence of the SU(2)$_\cC$ term in (\ref{projSuperWeyl})
which is due to the compensating SU(2)$_{\cC}$ transformations that appear in the super-Weyl 
transformation of the minimal supergravity covariant derivatives (\ref{SW-000}).

Let us also remind that, if $Q^{(m,n)}(z,u^\opl,v^\bpl)$ is a bi-projective multiplet,
its complex conjugate 
is not covariantly analytic.
However, one can introduce a generalized,  analyticity-preserving 
conjugation, $Q^{(m,n)} \to \widetilde{Q}^{(m,n)}$, defined as
\bsubeq
\bea
&&\widetilde{Q}^{(m,n)} (u^\opl,v^\bpl)\equiv \bar{Q}^{(m,n)}\big(
\overline{u^\opl}\to 
\widetilde{u}^\opl,
\overline{v^\bpl}\to 
\widetilde{v}^\bpl\big)~, 
\\
&&
\widetilde{u}^\opl = {\rm i}\, \s_2\, u^\opl~, 
~~~
\widetilde{v}^\bpl = {\rm i}\, \s_2\, v^\bpl~, 
\eea
\esubeq
with $\bar{Q}^{(m,n)}(\overline{u^\opl},\overline{v^\bpl})$ the complex conjugate of $Q^{(m,n)}$
and $\overline{u^\opl},\overline{v^\bpl}$ the complex conjugates of $u^\opl,v^\bpl$.
Then $\widetilde{Q}^{ (m,n) } (z,u^\opl,v^\bpl)$ 
is a  weight-(m,n) bi-projective multiplet.
One can see that
$\widetilde{\widetilde{Q}}{}^{(m,n)}=(-1)^{m+n}Q^{(m,n)}$,
and therefore real supermultiplets can be consistently defined when 
$(m+n)$ is even.

The simplest example of bi-projective superfield is given by the covariant 
twisted-II multiplet (TM-II) \cite{GTM-2DSG}.
Consider a superfield $T_{ij}$ satisfying a set of analyticity-like differential constraints
\cite{TM-II-IK} 
\bea
\de_{+ (k}T_{i)j}=\deb_{+(k}T_{i)j}=0~,~~~
\de_{-(k}T_{|i|j)}=\deb_{-(k}T_{|i|j)}=0
~.
\label{TMII-analyticity}
\eea
The superfield $T_{ij}$ is a Lorentz scalar and possesses  the SU(2)  transformations 
\bea
\bmL_{kl}T_{ij}=\hf C_{i(k}T_{l)j}~,~~~~~~
\bmR_{kl}T_{ij}=\hf C_{j(k}T_{|i|l)}~.
\label{LTRT}
\eea
Note that $T_{ij}$ has no symmetry conditions imposed in the $i$ and $j$ indices but 
satisfies the reality condition $(T_{ij})^*=T^{ij}$.

We have already seen an example of TM-II described by the super-Weyl transformation 
parameters $(\bS,\bS_{ij})$ constrained by (\ref{dbSbS}).
In fact, if one decomposes $T_{ij}$ in its symmetric and antisymmetric parts
 $T_{ij}=W_{ij}+\hf C_{ij}F$, 
where $W_{ij}=W_{ji}$ and both $W_{ij}$ and $F$ are real $(W_{ij})^*=W^{ij}$, $(F)^*=F$,
then the constraints (\ref{TMII-analyticity})  are equivalent to (\ref{dbSbS}) with $(F,W_{ij})$
taking the place of $(\bS,\bS_{ij})$.

By contracting the $u^\opl,\,v^\bpl$ 
isotwistors with $T_{ij}$, the superfield $T^{\opl\bpl}(z,u,v)$ is defined as
\bea
T^{\opl\bpl}(u,v):=u^\opl_i v^\bpl_jT^{ij}~.
\eea
Then, the analyticity conditions (\ref{TMII-analyticity}) are equivalent to (\ref{analyticity}). 
Moreover, the SU(2) transformations (\ref{LTRT}) can be written exactly as eqs. 
(\ref{LU})--(\ref{RU}) with $T^{\opl\bpl}$ considered as a weight-(1,1) isotwistor superfield.
Therefore, $T^{\opl\bpl}$ satisfies all the conditions of a weight-(1,1) bi-projective superfield.
By definition $T^{\opl\bpl}$ describes a regular holomorphic tensor field on the whole product 
of two complex projective spaces ${\mathbb C}P^1\times{\mathbb C}P^1$.
More general multiplets can have poles and more complicate analytic properties on
 ${\mathbb C}P^1\times{\mathbb C}P^1$.
 For instance one can easily define 2D bi-projective superfields with infinite number of 
superfields in a way completely analogue to the more studied curved 4D-5D cases
 \cite{KT-M_5D,KT-M_5Dconf,KLRT-M_4D-1,KLRT-M_4D-2}.
 The twisted-II multiplet play a special role also because it represent the conformal compensator 
 for the minimal supergravity.

The constraints of the covariant TM-II can be solved in terms of a prepotential described by a so 
called covariant twisted-I multiplet (TM-I) \cite{GTM-2DSG}.
The TM-I can be described by the superfields $W,\,P$ and $Q$ that 
are defined to be invariant under the action of the Lorentz $\cm$ and SU(2)s
$\bmL_{ij}$, $\bmR_{ij}$ generators.
Moreover the TM-I superfields are chosen to be invariant under 
super-Weyl transformations $\tilde{\d}W=\tilde{\d}P=\tilde{\d}Q=0$ and enjoy 
the following constraints\footnote{Note that the covariant TM-I constraints given here 
are equivalent  to the differential constraints (\ref{minimalCONSTR})
of the  torsion components $N,\,\cS$ and $\cT$ of
the minimal supergravity multiplet.
However, the two multiplet posses a crucial difference:
the superfields $(W,\,P,\,Q)$ are invariant under super-Weyl transformations while 
$(N,\,\cS,\,\cT)$
are not and transform inhomogeneously according to (\ref{minimalSW-1})--(\ref{minimalSW-3}).
This difference emphasize that, even if they consistently 
satisfy the same differential constraints,
$(W,\,P,\,Q)$ are matter superfields while $(N,\,\cS,\,\cT)$ are supregravity torsion components.}
\cite{Gnew2D,GHR,GatesKetov}
\bsubeq
\bea
&&\deb_\a^iW=0~,~~~
\de_{\g k}Q=\hf(\g^3)_\g{}^\d\deb_{\d k}\bar{W}~,~~~
\de_{\a i}P=-{\ri\over 2}\deb_{\a i}\bar{W}~,
\label{cov-TM-I-1}
\\
&&~~~~~~~~~~~~~~~~~~
(W)^*=\bar{W}~,~~~(P)^*=P~,~~~(Q)^*=Q~.
\eea
\esubeq
In (\ref{cov-TM-I-1}) we have omitted some constraints that can be obtained by complex 
conjugation.
The superfield $T^{\opl\bpl}$ of the TM-II can be then described in terms of a TM-I 
by the aid of the following equations \cite{GTM-2DSG}
\bea
T^{\opl\bpl}
=u^\opl_iv^\bpl_jT^{ij}
={\ri\over 4}u^\opl_iv^\bpl_j{[}\de_{+}^i,\de_{-}^j{]}W
={\ri\over 4}u^\opl_iv^\bpl_j{[}\deb_{+}^i,\deb_{-}^j{]}\bar{W}
=u^\opl_iv^\bpl_j(T_{ij})^*=\widetilde{(T^{\opl\bpl})}
~.~~~
\label{T-W}
\eea


We can now provide a bi-projective superfield action principle. 
This is invariant under the supergravity gauge 
group and super-Weyl transformations.
Let the Lagrangian $\cL^{(0,0)}$ be a real bi-projective superfield of weight-$(0,0)$.
Consider a TM-II described by $T^{\opl\bpl}$ with $W, (\bar{W})$ the chiral  
superfield of the TM-I prepotential.
Associated with $\cL^{(0,0)}$ we introduce the action principle
\bea
S={1\over 4 \pi^2}
 \oint (u^\opl \rd u^\opl)
  \oint (v^\bpl \rd v^\bpl)
\int \rd^2 x \,{\rm d}^8\q \,E\, \frac{W{\bar W}}{(T^{\opl\bpl})^2}\cL^{(0,0)}~, 
~~~ E^{-1}= {\rm Ber}(E_A{}^M)~.
~~~~~~
\label{InvarAc}
\eea
By construction, the functional is  invariant under the  re-scaling
$u_i^\opl(t)  \to c_L(t) \,u^\opl_i(t) $, for an arbitrary function
$ c_L(t) \in {\mathbb C}\setminus  \{0\}$, 
where $t$ denotes the evolution parameter 
along the first closed integration contour.
 Similarly, (\ref{InvarAc}) is invariant
under  re-scalings
$v_I^\bpl(s)  \to c_R(s) \,v^\bpl_I(s) $, for an arbitrary function
$ c_R(s) \in {\mathbb C}\setminus  \{0\}$, 
where $s$ denotes the evolution parameter 
along the second closed integration contour.
Note that (\ref{InvarAc}) has clear similarities with the action principles in four and five-dimensional
curved projective superspace  \cite{KT-M_5D,KT-M_5Dconf,KLRT-M_4D-1,KLRT-M_4D-2}.

The action (\ref{InvarAc}) can be proved to be invariant under arbitrary local supergravity gauge 
transformations (\ref{SUGRAgauge}). 
The invariance under general coordinates and Lorentz transformations is trivial.
One can prove the invariance under the two SU(2)$_L$ and SU(2)$_R$,
and then SU(2)$_\cV$ in (\ref{SUGRAgauge}),
 transformations.
By using that under super-Weyl transformations  $E$ varies like $\tilde{\d}E=2\bS E$
and the transformations 
$\tilde{\d}\cL^{(0,0)}=-\bS^{kl}\cC_{kl}\cL^{(0,0)}$,
$\tilde{\d}T^{\opl\bpl}=(\bS-\bS^{kl}\cC_{kl})T^{\opl\bpl}$ and $\tilde{\d}W=W$, 
one sees that $S$ is super-Weyl invariant. 
Moreover, it is important to note that one can prove \cite{GTM-2DSG} that if $\cL^{(0,0)}$ 
is a function of some 
supermultiplets to which the TM-II compensator does not belong, then the 
action $S$ is independent of the superfields  $T^{\opl\bpl}$, $W$ and $\bar{W}$ chosen.

It would be clearly of interest to reduce the bi-projective action principle  (\ref{InvarAc}) 
to components and find the bi-projective density operator analogously to the chiral action
of section \ref{ecto}. One could derive the action (\ref{InvarAc}) in components
by using the ``projective-invariance'' techniques similarly to
the 5D $\cN=1$ \cite{KT-M_5D} and 4D $\cN=2$ \cite{KT-M-normal}  cases.
Alternatively, and more interestingly, one could use ectoplasm \cite{Gates,GGKS,GKT-M_2009}
or normal coordinates techniques \cite{GKS,KT-Mnc}

\section{A bi-projective prepotential for the covariant TM-I}
\label{TM-I_prep}

This section is devoted to present some new results on the 
bi-projective superspace formalism of \cite{GTM-2DSG}.
In particular, here we give the solution of the covariant twisted-I multiplet constraints
(\ref{cov-TM-I-1}) in terms of a weight-$(-1,-1)$ real but otherwise unconstrained
bi-projective superfield $V^{(-1,-1)}$.
Although in this paper for simplicity 
we are focusing on the minimal supergravity described in section 
\ref{SUGRA}, it is important to point out that all the results in this section 
remain true without any modifications if one consider the extended SU(2)$_L\times$SU(2)$_R$ 
superspace supergravity geometry of \cite{GTM-2DSG}.

Let us start by giving the result.
Consider the superfields
\bsubeq
\bea
W&=&
{1\over 4\pi^2}
 \oint {(u^\opl \rd u^\opl)\over(u^\opl u^\omn)}
  \oint {(v^\bpl \rd v^\bpl)\over(v^\bpl v^\bmn)}
  \deb_+^{\omn}\deb_-^{\bmn}V^{(-1,-1)}
  ~,
\label{W-V}
  \\
X&=&
-{1\over 4\pi^2}
 \oint {(u^\opl \rd u^\opl)\over(u^\opl u^\omn)}
  \oint {(v^\bpl \rd v^\bpl)\over(v^\bpl v^\bmn)}
  \deb_+^{\omn}\de_-^{\bmn}V^{(-1,-1)}
  ~,
  \label{X-V}
\eea
\esubeq
these turn out to describe a covariant twisted-I multiplet
where the superfields $P$ and $Q$ have been reabsorbed into the complex superfield $X$ 
defined as
\bea
X=Q+\ri P
~,~~~~~~
\bar{X}=(X)^*
~.
\label{defX}
\eea
According to (\ref{W-V})--(\ref{X-V}), and provided that $V^{(-1,-1)}$ is 
a weight-$(-1,-1)$ bi-projective superfield,
the $W$ and $X$ superfields 
are invariant under Lorentz, SU(2)$_L$, SU(2)$_R$ and super-Weyl 
transformations.
Moreover, they satisfy the following differential constraints
\bsubeq
\bea
&&\deb_{+}^iW=0~,~~
\deb_{+}^{i}X=0~,~~
\deb_{+}^{i}\bar{X}=-\de_{+}^{i}W~,
~~~~
\label{cov-TM-I-1-001}
\\
&&\deb_{-}^{i}W=0~,~~
\de_{- i}X=0~,~~~
\deb_{-}^{i}X=\de_{-}^{i}W~.
~~~~
\label{cov-TM-I-1-002}
\eea
\esubeq
The previous equations, once used (\ref{defX}),
are indeed equivalent to (\ref{cov-TM-I-1}).

Let us provide some details of the proof that, as stated above, $W$ and $X$ defined in terms
of $V^{(-1,-1)}$ satisfy all the properties of the covariant TM-I.

First, let us note that it holds 
\bea
 \oint {(u^\opl \rd u^\opl)\over(u^\opl u^\omn)}
  \oint {(v^\bpl \rd v^\bpl)\over(v^\bpl v^\bmn)}
  \{\deb_+^{\omn},\deb_-^{\bmn}\}V^{(-1,-1)}=0
  ~.
  \label{D-D-V--}
 \eea
Analogously, the integral 
$ \oint {(u^\opl \rd u^\opl)\over(u^\opl u^\omn)}
  \oint {(v^\bpl \rd v^\bpl)\over(v^\bpl v^\bmn)}
\{ \deb_+^{\omn},\de_-^{\bmn}\}V^{(-1,-1)}$
is also zero.
Then, one can freely anticommute the derivatives, and consider only the commutator part
in the equations (\ref{W-V})--(\ref{X-V}).
The equation (\ref{D-D-V--}) can be proved by using the minimal covariant derivatives algebra,
the following relation 
 \bea
 \cV^{\omn\bmn}V^{(-1,-1)}&=& 
-D^{\omn\omn} (u^\opl v^{\bmn}) V^{(-1,-1)}
-D^{\bmn\bmn} (v^\bpl u^\omn)V^{(-1,-1)}
~,
\eea
which easily follows from (\ref{LU})--(\ref{RU}), and by using the fact that 
it  holds
\bea
\oint {(u^\opl \rd u^\opl)\over(u^\opl u^\omn)}D^{\omn\omn}f_L^{(0)}(u^\opl)=0~,~~~~~~
  \oint {(v^\bpl \rd v^\bpl)\over(v^\bpl v^\bmn)}D^{\bmn\bmn}f_R^{(0)}(v^\bpl)=0
  \label{intD--}
\eea
for any function $f_L^{(0)}(u^\opl)$ homogeneous of degree zero in $u^\opl$
and any function $f_R^{(0)}(v^\bpl)$ homogeneous of degree zero in $v^\bpl$.

It is important to note that $W$ and ${X}$ 
do not depend on the isotwistors $u^\omn$ and $v^\bmn$ even if they explicitly appear 
on the right hand side of (\ref{W-V})--(\ref{X-V}).
In particular,  (\ref{W-V})--(\ref{X-V})
are invariant under arbitrary ``projective'' transformations of the form:
\bea
(u_i^\omn\,,\,u_i^\opl)~\to~(u_i^\omn\,,\, u_i^\opl )\,P_L~,~~~~~~P_L\,=\,
\left(\begin{array}{cc}a_L~&0\\ b_L~&c_L~\end{array}\right)\,\in\,{\rm GL(2,\mathbb{C})}~,
\label{projectiveGaugeVar-L}
\\
(v_i^\bmn\,,\,v_i^\bpl)~\to~(v_i^\bmn\,,\, v_i^\bpl )\,P_R~,~~~~~~P_R\,=\,
\left(\begin{array}{cc}a_R~&0\\ b_R~&c_R~\end{array}\right)\,\in\,{\rm GL(2,\mathbb{C})}~.
\label{projectiveGaugeVar-R}
\eea
These transformations express the homogeneity  with respect to
 $u^\opl,\,v^\bpl$ and the independence on $u^\omn,\,v^\bmn$. 
The invariance of  (\ref{W-V})--(\ref{X-V})
 under the $a$ and $c$ part of the transformations are trivial. 
Let us see that it is true also for $b$-transformations. For example, consider 
$\d_{b_R} v^\omn=b_R v^\opl$ in (\ref{W-V})
\bea
\d_{b_R} W=
{1\over 4\pi^2}
 \oint {(u^\opl \rd u^\opl)\over(u^\opl u^\omn)}
  \oint {(v^\bpl \rd v^\bpl)\over(v^\bpl v^\bmn)}
b_R  \deb_+^{\omn}\deb_-^{\bpl}V^{(-1,-1)}
=0
  ~,
\eea
which is zero being $V^{(-1,-1)}$ a bi-projective superfield.
By using (\ref{D-D-V--}) and then considering the $b_L$-transformation 
it similarly holds $\d_{b_L}W=0$. Analogously, it follows the invariance under 
(\ref{projectiveGaugeVar-L})--(\ref{projectiveGaugeVar-R}) of the right hand side of
(\ref{X-V}).

The Lorentz invariance of (\ref{W-V})--(\ref{X-V}) is trivial.
Let us prove the SU(2) invariance.
By using (\ref{RU}), (\ref{R_kl}) and then (\ref{intD--}),  one can prove 
\bea
\bmR_{kl} W=
-{1\over 4\pi^2}
 \oint {(u^\opl \rd u^\opl)\over(u^\opl u^\omn)}
  \oint {(v^\bpl \rd v^\bpl)\over(v^\bpl v^\bmn)}
D^{\bmn\bmn}{v^\bpl_{(k}v^{\bpl}_{l)}\over 2(v^\bpl v^\bmn)}  \deb_+^{\omn}\deb_-^{\bmn}
V^{(-1,-1)}
=0
~.
\eea
In a very similar manner one obtains that $W$ and $X$ 
under the action of $\bmL_{kl}$ and $\bmR_{kl}$.

By using the equations (\ref{SW-000}), (\ref{L_kl}), (\ref{R_kl}), (\ref{LU})--(\ref{LorentzU}), 
(\ref{analyticity}) and (\ref{intD--}),
it is not difficult to prove that $W$ and ${X}$ in (\ref{W-V})--(\ref{X-V})
are invariant under super-Weyl transformations. We leave this computation to the reader.

To prove that $W$ and ${X}$ describe a covariant TM-I it is left to
prove that (\ref{cov-TM-I-1-001})--(\ref{cov-TM-I-1-002}) are satisfied.
Let us prove that $W$ in (\ref{W-V}) is chiral. 
By using 
\bea
(u^\opl u^\omn)\d_{j}^i=(u^{\opl i}u^\omn_j-u^\opl_j u^{\omn i} )
~,
\eea
and the analyticity of $V^{(-1,-1)}$, we find
\bea
\deb_+^iW&=&
{1\over 4\pi^2} \oint {(u^\opl \rd u^\opl)\over(u^\opl u^\omn)^2}
  \oint {(v^\bpl \rd v^\bpl)\over(v^\bpl v^\bmn)}
\Big(
\hf u^{\opl i}\{\deb_+^\omn,  \deb_+^{\omn}\}\deb_-^{\bmn}
-u^{\omn i}\{\deb_+^\opl,  \deb_+^{\omn}\}\deb_-^{\bmn}
\non\\
&&~~~~~~~~~~~~~~~~~~~~~~~~~~~~~~~~~~~~~~~
+u^{\omn i}\deb_+^{\omn}\{\deb_+^\opl,\deb_-^{\bmn}\}
\Big)V^{(-1,-1)}
~.
\label{DbW-1}
\eea
By considering that it holds
\bea
\{\deb_+^i,\deb_+^j\}=0
~,~~~~~~
\{\deb_{+}^{i},\deb_{-}^{j}\}=
-4C^{ij}\bar{T}\cM
-4\bar{T}\cV^{ij}
~,
\eea
and
\bea
\cV^{\opl\bmn}V^{(-1,-1)}&=&
D^{\bmn\bmn} (u^\opl v^\bpl)V^{(-1,-1)}
~,
\eea
which follows from (\ref{LU})--(\ref{RU}), one easily obtains
\bea
\deb_{+}^iW=0~.
\eea
Similarly one finds that $\deb_{-}^iW=0$ and then $\deb_{\a}^iW=0$.
Analogously, it can be derived $\deb_+^iX=0$ and $\de_{-i}{X}=0$.

Let us now turn our attention to the equation $\deb_{+ i}\bar{X}=-\de_{+ i}W$ in 
(\ref{cov-TM-I-1-001}).
We obtain
\bea
\de_{+i}W&=&
{1\over 4\pi^2} 
 \oint {(u^\opl \rd u^\opl)\over(u^\opl u^\omn)^2}
  \oint {(v^\bpl \rd v^\bpl)\over(v^\bpl v^\bmn)}
\Big(
u^\opl_i\de_{+}^\omn  \deb_+^{\omn}\deb_-^{\bmn}
-u^\omn_i\{\de_{+}^\opl,  \deb_+^{\omn}\}\deb_-^{\bmn}
\non\\
&&~~~~~~~~~~~~~~~~~~~~~~~~~~~~~~~~~~~~~~~
+u^\omn_i  \deb_+^{\omn}\{\de_{+}^\opl,\deb_-^{\bmn}\}
\Big)V^{(-1,-1)}
~.
\eea
By using the minimal supergravity anticommutators
\bea
\{\de_{+}^i,\deb_+^j\}=
2\ri C^{ij}\de_{++}
~,~~~
\{\de_{+ i},\deb_{-j}\}=
4\ri C_{ij}\cS\cM
-4C_{ij}\cT\cM
+4\cT\cV_{ij}
-4\ri\cS\cV_{ij}
~,~~~
\eea
it follows
\bea
\de_{+i}W=
{1\over 4\pi^2}
 \oint {(u^\opl \rd u^\opl)\over(u^\opl u^\omn)}
  \oint {(v^\bpl \rd v^\bpl)\over(v^\bpl v^\bmn)}
\Bigg(
{u^\opl_i\over(u^\opl u^\omn)}\de_{+}^\omn  \deb_+^{\omn}\deb_-^{\bmn}
+2\ri u^\omn_i\de_{++}\deb_-^{\bmn}
\Bigg)V^{(-1,-1)}
~.~~~
\eea
Next, we compute $\deb_{+ i}\bar{X}$
\bea
\deb_{+i}\bar{X}&=&
{1\over 4\pi^2}
 \oint {(u^\opl \rd u^\opl)\over(u^\opl u^\omn)^2}
  \oint {(v^\bpl \rd v^\bpl)\over(v^\bpl v^\bmn)}
\Big(
-u^\opl_i \de_+^{\omn}\deb_{+}^\omn \deb_-^{\bmn}
+u^\opl_i\{  \de_+^{\omn},\deb_{+}^\omn\}\deb_-^{\bmn}
\non\\
&&~~~~~~~~~~~~~~~~~~
-u^\omn_i\{\de_{+}^\omn,  \deb_+^{\opl}\}\deb_-^{\bmn}
+u^\omn_i  \de_+^{\omn}\{\deb_{+}^\opl,\deb_-^{\bmn}\}
\Big)V^{(-1,-1)}
~.
\eea
One can simplify the previous equation and prove that
 $\deb_{+}^i\bar{X}=-\de_{+}^iW$. Analogous computations
can be used to derive $\deb_-^iX=\de_-^iW$. 
Then, $W$ and $X$ in eqs. 
(\ref{W-V})--(\ref{X-V}) satisfy all the defining properties of the covariant type-I twisted multiplet. 

Let us conclude by giving some comments about the results of this section.
First of all, the prepotential solution of the TM-I has clear analogies with the 
weight-zero real projective prepotential of the 
chiral field strength of an Abelian vector multiplet in 4D $\cN=2$ superspace supergravity 
\cite{KLRT-M_4D-1,KLRT-M_4D-2}.
We remind that in the 4D case, the projective prepotential $V^{(0)}$ 
possesses the gauge freedom $\d_gV^{(0)}=\L^{(0)}+\widetilde{\L}^{(0)}$
where $\L^{(0)}$ is a weight-zero covariantly arctic superfield and $\widetilde{\L}^{(0)}$ is its 
analyticity preserved conjugate.
The 4D arctic superfield  $\L^{(0)}$ is such that in the north chart of ${\mathbb C}P^1$
it does not possess any poles.
In our 2D bi-projective case  the solution (\ref{W-V})--(\ref{X-V}) turns out to
possess a gauge freedom
\bea
\d_gV^{(-1,-1)}=
\L_L^{(-1,-1)}+\widetilde{\L}_L^{(-1,-1)}
+\L_R^{(-1,-1)}+\widetilde{\L}_R^{(-1,-1)}
~,
\label{gaugeV}
\eea
 generated by the superfields
$\L_L^{(-1-1)}(u^\opl,v^\bpl)$ and $\L_R^{(-1-1)}(u^\opl,v^\bpl)$ together with their conjugated.
Here the superfields 
$\L_L^{(-1-1)}$ and $\L_R^{(-1-1)}$,
are such that $\L_L^{(-1-1)}$ does not possess poles on the north chart of the left
${\mathbb C}P^1$ having homogeneous coordinates $u^\opl$ and
$\L_L^{(-1-1)}$ does not possess poles on the north chart of the right
${\mathbb C}P^1$ having homogeneous coordinates $v^\bpl$.
In this way either the $u^\opl$ or the $v^\bpl$ contour integral in the definition of the field strengths
(\ref{W-V})--(\ref{X-V}) is zero.

Considering that the covariant TM-I describe a prepotential for the TM-II,
It is clear that one can solve the type-II twisted multiplet constraints in terms of $V^{(-1,-1)}$
by using the equations (\ref{W-V}) and (\ref{T-W}).\footnote{It is worth noting that in the flat case a 
similar prepotential solution of the TM-II constraints has been described by Siegel in 
\cite{Siegel2DN4} by using a form of bi-projective superspace.}
Now, given a TM-II described by the superfield
$\bT^{\opl\bpl}$ and its projective prepotential $\bV^{(-1,-1)}$ one can construct
an action by considering the bi-projective Lagrangian\footnote{Here the TM-II and its prepotential 
do not have to be the supergravity conformal compensator;
this is why we have used the bold characters to distinguish it by the one in (\ref{InvarAc}).}
\bea
{\bf{\cL}}^{(0,0)}=\bV^{(-1,-1)}\bT^{\opl\bpl}
~.
\label{VT}
\eea 
The action (\ref{InvarAc}),  with the previous Lagrangian, is then
 invariant under (\ref{gaugeV}).

To conclude we stress again that, if one considers the 
SO(1,1)$\times$SU(2)$_{L}\times$SU(2)$_R$ 
extended supergravity geometry of \cite{GTM-2DSG}
all the main results in this section, in particular 
(\ref{W-V})--(\ref{X-V}), remain unchanged even if the computations described here become
a bit more complicate.

~~~\\
\noindent
{\bf Acknowledgements:}\\
We are grateful to S. James Gates, Jr. 
to have raised our interest on 2D $\cN=(4,4)$ superspace supergravity 
and for the joint collaboration in \cite{GatesGT-M-09-1}.
We thank the organizers of the Workshop ``Supersymmetries and Quantum Symmetries''
(SQSÕ09), at the Bogoliubov Laboratory of Theoretical Physics, JINR, Dubna, 
July 29--August 3 2009,
for the opportunity to report on results of the research reviewed in this contribution.
This research was supported  by the endowment 
 of the John S.~Toll Professorship, the University of 
 Maryland Center for String \& Particle Theory, and
 National Science Foundation Grant PHY-0354401.


\begin{small}

\end{small}

\end{document}